\documentclass{article}
\usepackage{graphicx}
\usepackage{amssymb}
\usepackage[compress]{cite}
\usepackage{dcolumn}% Align table columns on decimal point
\usepackage{color}
\usepackage{epstopdf}
\usepackage{amsmath}
\usepackage{amsfonts}
\usepackage{mathrsfs}
\usepackage{varioref}
\usepackage[active]{srcltx}
\usepackage{tikz}
\usepackage{textcomp}
\usepackage{tikz,pgfplots}
\usepackage{lmodern}
\usepackage[caption=false]{subfig}
\usepackage{paralist}

\usepackage{ccfonts}
\usepackage{multicol}
\usepackage[english]{babel}
\usepackage{amsmath}
\usepackage{multirow}
\usepackage{graphics}
\usepackage{colortbl}
\usepackage{colordvi}
\usepackage{verbatim}
\usepackage{mathtools}
\usepackage{braket}
\usepackage{tikz}
\usepackage{amssymb}
\usepackage{slashed}
\usepackage{color,soul}
\setul{0.5ex}{0.3ex}
\setulcolor{red}
\usepackage{diagbox}
\usepackage{newcent}

\usetikzlibrary{decorations.pathmorphing}
\tikzset{snake it/.style={decorate, decoration=snake}}
\def\KR{Kalb-Ramond }

\def\gr{general relativity}
\newcommand{\bea}{\begin{eqnarray}}
\newcommand{\eea}{\end{eqnarray}}
\def\gr{general relativity}

\def\a {\alpha}
\def\b{\beta}
\def\m{\mu}
\def\n{\nu}
\def\g{\gamma}

\RequirePackage[colorlinks,citecolor=blue,urlcolor=magenta,linkcolor=blue]{hyperref}
\def\gr{general relativity}

\def\KR{Kalb-Ramond }
\def\a {\alpha}
\def\b{\beta}
\def\m{\mu}
\def\n{\nu}
\def\g{\gamma}
\labelformat{section}{Section #1} 
\labelformat{subsection}{Section #1} 
\labelformat{subsubsection}{Section #1}
\labelformat{subsubsubsection}{Section #1}
\labelformat{equation}{Eq.~(#1)} 
\labelformat{figure}{Fig.~#1} 
\labelformat{subfigure}{Fig.~\thefigure#1} 
\labelformat{table}{Tab.~#1} 
\labelformat{appendix}{Appendix #1}
\title{Implications of axionic hair on shadow of M87*}
\author{Indrani Banerjee
\footnote{tpib@iacs.res.in},
Subhadip Sau
\footnote{tpss2@iacs.res.in},
and Soumitra SenGupta
\footnote{tpssg@iacs.res.in}\\
{\small{School of Physical Sciences, Indian Association for the Cultivation of Science, Kolkata-700032, India}}}

\date{ }  %% This command  will supress printing the date. 
\begin{document} 
\maketitle
%%%%%%%%%%%%%%%%%%%%%%%%%%%%%%%%%%%%%%%%%%%%%%%%%%%%%%%%%%%%%%%%%%%%%%%%%%%%%%%%%%%%%%%%%%%%%%%%%%%
%%%%%%%%%%%%%%%%%%%%%%%%%%%%%%%%%%%%%%%%%%%%%%%%%%%%%%%%%%%%%%%%%%%%%%%%%%%%%%%%%%%%%%%%%%%%%%%%%%%
%%%%%%%%%%%%%%%%%%%%%%%%%%%%%%%%%%%%%%%%%%%%%%%%%%%%%%%%%%%%%%%%%%%%%%%%%%%%%%%%%%%%%%%%%%%%%%%%%%%
\begin{abstract}
Detection of axion field can unfold intriguing facets of our universe in several astrophysical and cosmological scenarios. In four dimensions, such a field owes its origin to the completely anti-symmetric Kalb-Ramond field strength tensor.
Its invisibility in the solar system based tests compels one to look for its signatures in the strong field regime. The recent observation of the shadow of the supermassive black hole in the galaxy M87 ushers in a new opportunity to test for the footprints of axion in the near horizon region of black holes, where the gravity is expected to be strong. In this paper, we explore the impact of axion on the black hole shadow and compare the result with the available image of M87*. Our analysis indicates that axion which violates the energy condition seems to be favored by observations. The implications are discussed.    

%The absence of such a field in the solar system based tests compels one to look for its signatures in the strong field regime. The recent observation of the shadow of the supermassive black hole in the galaxy M87 ushers in a new opportunity to test for the footprints of such a field in near horizon regime of black holes.

\end{abstract}

%%%%%%%%%%%%%%%%%%%%%%%%%%%%%%%%%%%%%%%%%%%%%%%%%%%%%%%%%%%%%%%%%%%%%%%%%%%%%%%%%%%%%%%%%%%%%%%%%%%
%%%%%%%%%%%%%%%%%%%%%%%%%%%%%%%%%%%%%%%%%%%%%%%%%%%%%%%%%%%%%%%%%%%%%%%%%%%%%%%%%%%%%%%%%%%%%%%%%%%
%%%%%%%%%%%%%%%%%%%%%%%%%%%%%%%%%%%%%%%%%%%%%%%%%%%%%%%%%%%%%%%%%%%%%%%%%%%%%%%%%%%%%%%%%%%%%%%%%%%
%\newpage
%%%%%%%%%%%%%%%%%%%%%%%%%%%%%%%%%%%%%%%%%%%%%%%%%%%%%%%%%%%%%%%%%%%%%%%%%%%%%%%%%%%%%%%%%%%%%%%%%%%
%%%%%%%%%%%%%%%%%%%%%%%%%%%%%%%%%%%%%%%%%%%%%%%%%%%%%%%%%%%%%%%%%%%%%%%%%%%%%%%%%%%%%%%%%%%%%%%%%%%
%%%%%%%%%%%%%%%%%%%%%%%%%%%%%%%%%%%%%%%%%%%%%%%%%%%%%%%%%%%%%%%%%%%%%%%%%%%%%%%%%%%%%%%%%%%%%%%%%%%
\section{Introduction}
Axions are pseudo-scalar fields which appear as closed string excitations in the heterotic string spectrum \cite{Kalb:1974yc,Green:1987mn}. In four dimensions,
the derivative of such a field is associated with the Hodge dual of the Kalb-Ramond field strength $H_{\a\m\n}$, which plays a significant role in explaining several astrophysical and cosmological observations. The field strength tensor $H_{\a\m\n}$ transforms like a third rank completely anti-symmetric tensor field and is associated with a massless, second rank anti-symmetric tensor $B_{\mu \nu}$, the so called Kalb-Ramond field. In higher-dimensional theories such a field is necessary to unify gravity and electromagnetism \cite{doi:10.1063/1.530877,German:1993bq}. 
Apart from the emergence of such 3-forms in the effective low energy action of a type IIB string theory \cite{Kalb:1974yc,Green:1987mn}, they play consequential roles in understanding leptogenesis \cite{Mavromatos:2012cc,ELLIS2013407}, in explaining the cosmic microwave background anisotropy \cite{Lue:1998mq,1475-7516-2004-06-005} and in engendering topological defects which are instrumental in imparting intrinsic angular momentum to galaxies \cite{Chandia:1997hu,0264-9381-12-2-016}. 

The emergence of superstring theory \cite{Kalb:1974yc,Green:1987mn} provided a further incentive to investigate the 
nature and consequences of the Kalb-Ramond field. Its compelling resemblance with space-time torsion \cite{Majumdar:1999jd,PhysRevD.81.024021,Chakraborty:2017uku,Sur:2016bwu,0264-9381-12-2-016,Hehl:1976kj,deSabbata:1994wi,Capozziello:2001mq,SenGupta:2001yj,Lue:1998mq} is noteworthy. In general relativity, the third rank torsion tensor $T^\alpha_{\mu\nu}$ is associated with the anti-symmetric part of the affine connection, i.e., $T^\alpha_{\mu\nu}=\Gamma^\alpha_{\mu\nu}-\Gamma^\alpha_{\nu\mu}$ and is primarily anti-symmetric in two indices. Its association with the Kalb-Ramond field strength $H_{\a\m\n}$ becomes evident only when we consider a special sub-class of the  torsion tensor antisymmetrized in all the three indices \cite{Majumdar:1999jd,Hehl:1976kj,deSabbata:1994wi,Sur:2016bwu}. In such a scenario, Einstein gravity with the Kalb-Ramond field in the matter sector is equivalent to a modified theory of gravity incorporating the completely anti-symmetric space-time torsion. Due to this remarkable analogy between spacetime torsion and Kalb-Ramond field, gravity theories based on twistors necessitates Kalb-Ramond field \cite{HOWE199680} and one can show that such a field can successfully generate  optical activity in spacetime exhibiting birefringence \cite{0264-9381-19-4-304,Kar:2000ct}.

%A further impetus to explore the theoretical and observational consequences of Kalb-Ramond field arises from the inefficacy of \gr\ in adequately addressing the dark sector. Observations related galactic rotation curves and accelerated expansion of the universe indicate   \cite{CLIFTON20121,0004-637X-517-2-565,1538-3881-116-3-1009} the need for either some additional matter fields or some alteration in the gravity sector or both. 
Moreover, the inefficacy of \gr\ in adequately addressing the dark sector \cite{Milgrom:1983pn,CLIFTON20121,0004-637X-517-2-565,1538-3881-116-3-1009} indicates the need for either some additional matter fields or some alteration in the gravity sector or both \cite{Ishak:2018his}. In such a scenario, inclusion of axions in the matter sector is often considered \cite{Dashko:2018dsw}.  
Although working with Kalb-Ramond field or completely
antisymmetric spacetime torsion corresponds to the same physical scenario, in this work we will concentrate primarily on modification in the matter sector due to the addition of the Kalb-Ramond field. Given the theoretical significance of such a field, it is instructive to search for the signatures of Kalb-Ramond field or axions in the available astrophysical and cosmological observations. The attempts to detect the presence of axion in solar system based tests, e.g. bending of light, perihilion precession of Mercury etc., reveal that such fields cause minuscle changes compared to \gr\ and hence cannot be detected by the present level of precision in the solar system based tests\cite{Kar:2002xa}. A quest for such a field in the spectrum of quasars have surprisingly revealed that axions which violate the energy condition seem to be favored by 
astrophysical observations related to black hole accretion \cite{Banerjee:2017npv}.
Incidentally, the observed spectrum of the same quasars seem to favor certain classes of alternate gravity theories, e.g., extra dimensions, Einstein Gauss-Bonnet gravity in higher dimensions, etc. \cite{Banerjee:2017hzw,Banerjee:2019sae,Banerjee:2019cjk}.

The recent observation of the shadow of the supermassive black hole M87* by the Event Horizon Telescope Collaboration \cite{Akiyama:2019cqa,Akiyama:2019brx,Akiyama:2019sww,Akiyama:2019bqs,Akiyama:2019fyp,Akiyama:2019eap} has enabled \emph{direct} observations of the near horizon regime of a black hole. This has opened up a new and independent window to test the nature of strong gravity. The high spatial resolution of the Event Horizon Telescope has facilitated polarimetric imaging of supermassive black holes like M87* which can be a possible probe to detect the presence of axionic particles \cite{Chen:2019fsq}. Moreover, based on the findings of the Event Horizon Telescope Collaboration, efforts are being made to establish constraints on the mass of ultralight scalar and vector bosons which can act as potential dark matter candidates \cite{Davoudiasl:2019nlo}.

The aim of this paper is therefore to examine the implications and consequences of axions/Kalb-Ramond field from the observed shadow of M87* which will enable us to understand whether the silhouette of M87* favors the presence of such a field.

\par
The paper is broadly classified into five sections. In \ref{S2}, we study the Einstein field equations with Kalb-Ramond field as the source and revisit the static, spherically symmetric and asymptotically flat black hole solution of such equations. \ref{S3} is dedicated in investigating the nature of the black hole shadow first in a general spherically symmetric background in \ref{S3a} and subsequently in \ref{S3b} we specialize to the spacetime with axionic hairs presented in \ref{S2}. In \ref{S4} we investigate the 
consequences of the Kalb-Ramond field on the recent observation of the shadow of M87*, the supermassive black hole located at the centre of the galaxy M87 and finally we conclude with a summary of our findings and implications of our results in \ref{S5}. 

Throughout the paper, the gravitational constant $G$ and the speed of light $c$ are taken to be unity. The metric convention adopted is (-,+,+,+,).

\section{Static spherically symmetric black hole solution in presence of Kalb-Ramond field}\label{S2}
In this section we discuss the nature of static, spherically symmetric black hole solution in presence of Kalb-Ramond field minimally coupled to gravity \cite{Kar:2002xa,SenGupta:2001cs}. 
The Kalb-Ramond field $B_{\m\n}$, which transforms like a second rank skew-symmetric tensor field can be considered to be a generalization of the electromagnetic four potential $A_\mu$ \cite{Kalb:1974yc,Majumdar:1999jd}. 
The field strength tensor $H_{\a\m\n}$ associated with the field $B_{\m\n}$ consists of a third-rank anti-symmetric tensor field and is given by,
\begin{align} 
H_{\a\m\n}=\partial_{[\a}B_{\m\n]}=\dfrac{1}{3}\left[\nabla_{\a}B_{\m\n}+\nabla_{\m}B_{\n\a}+\nabla_{\n}B_{\a\m}\right]=\dfrac{1}{3}\left[\partial_{\a}B_{\m\n}+\partial_{\m}B_{\n\a}+\partial_{\n}B_{\a\m}\right]
\label{1}
\end{align} 
The action associated with the Kalb-Ramond field in four dimensional Einstein gravity is given by
\begin{align}
S=\int \,d^{4}x\, \sqrt{-g}\left[\dfrac{R}{2\kappa^2}-\dfrac{1}{12} H_{\a\m\n}H^{\a\m\n}\right]
\label{2}
\end{align}
where, $g$ is the determinant of the metric tensor, $R$ is the Ricci Scalar and $\kappa=\sqrt{8\pi G}$, is related to the four dimensional gravitational constant $G$. 
The factor of $-{1}/{12}$ has been introduced in the Lagrangian so that one can have the canonical kinetic term as ${1}/{2}(\partial_{t}B_{\m\n})^2$ in the local inertial frame.
Field equations for \KR field can be derived by varying the action \ref{2} with respect to the field $B_{\m\n}$ which yields $\nabla_{\m}H^{\m\n\rho}=0$ as the equations of motion. By inspecting the equations of motion it can be shown that only the spatial components of the field are dynamical. This reduces the propagating degrees of freedom of this field to three, although the Kalb-Ramond field $B_{\m\n}$ possesses six independent components in four dimensions.  
A gauge symmetry $B_{\m\n}\rightarrow B_{\m\n}+\nabla_{[\m}\chi_{\n]}$ further reduces the degrees of freedom to zero as the gauge field $\chi_{\m}$ has three spatial components. However, the gauge field $\chi_{\m}$ exhibits a further invariance $\chi_{\m} \rightarrow \chi_{\m}+\partial_{\m} \Phi$, where $\Phi$ is a scalar field and in fact this is the scalar propagating degree of freedom for the \KR field in four dimensions.
Additionally, it can be shown that the \KR field satisfies the Bianchi identity given by $\nabla_{[\m}H_{\a\b\g]}=0$. For a more detailed discussion on the degrees of freedom of the \KR field in arbitrary dimensions, one is referred to \cite{Chakraborty:2017uku}.
Since the \KR field has a single propagating degree of freedom in four dimensions one can express its field strength $H_{\a\b\m}$ (which is a third rank anti-symmetric tensor field) in terms of the Hodge dual of the derivative of a pseudo-scalar field $\Phi$, known as the axion, where
 \begin{align}\label{5}
 H^{\m\n\rho}=\epsilon^{\m\n\rho\sigma}\partial_{\sigma}\Phi
 \end{align}
\ref{5} enables us to establish the connection between the Kalb-Ramond field with the axion and throughout the paper the terms axion and Kalb-Ramond field will be synonymously used.

The variation of the action \ref{2} with respect to the metric $g_{\mu\nu}$ leads to the gravitational field equations
\begin{align}\label{3}
G_{\m\n}=8\pi G\,\, T_{\m\n}^{(KR)}
\end{align}
where, $G_{\mu\nu}$ is the Einstein tensor and $T_{\m\n}^{(KR)}$ is the energy-momentum tensor for the \KR field given by
\begin{flalign}\label{4}
 T_{\m\n}^{(KR)}&= -\dfrac{2}{\sqrt{-g}}\dfrac{\delta(\sqrt{-g}\mathcal{\tilde{L}})}{\delta g^{\m\n}}\nonumber \\
&= \dfrac{1}{6}\left[3H_{\m\rho \sigma} H^{\,\,\rho \sigma}_{\n}-\dfrac{1}{2}g_{\m\n}\left(H_{\rho \sigma \delta} H^{\rho \sigma \delta}\right)\right]
\end{flalign}
such that $\mathcal{\tilde{L}}$ is the Lagrangian for the \KR field.

Since our goal in this paper is to explore the impact of axions on the shadow of the black hole, we first need to derive the static, spherically symmetric and asymptotically flat black hole solution of the Einstien's equations given in \ref{3}. This enables us to consider a line element of the form

 \begin{align}\label{6}
 ds^2= -e^{\n({r})} \,dt^2 +e^{\lambda({r})}\, d{r}^2 +{r}^{2}d\Omega^2
 \end{align}
 such that the metric elements $e^{\n({r})}$ and $e^{\lambda({r})}$ satisfying the Einstein's equations turn out to be two infinite series in $1/{r}$ \cite{Kar:2002xa,SenGupta:2001cs},
 \begin{align}\label{6a}
 e^{\n({r})}=1-\dfrac{2M}{{r}}+\dfrac{{b} M}{{r}^3}+\frac{2{b}M^2}{{r}^4} +\frac{72{b}M^3-27{b}^2 M}{20{r}^5}+ ...
 \tag{6a} 
 \end{align}
 \begin{align}\label{6b}
 e^{-\lambda({r})}=1-\dfrac{2M}{{r}}+\dfrac{3{b}}{{r}^2}+\frac{3{b}M}{{r}^3} +\frac{4{b}M^2}{{r}^4}+\frac{6M^3{b}}{{r}^5}-\frac{3{b}^2 M}{4{r}^5}+ ...
 \tag{6b}
 \end{align}
% \end{subequations}
where $r$ represents the radial distance from the black hole and $b$ is the axion parameter having units of $M^2$ (where we have assumed $G=c=1$). In what follows, we will scale $b$ by $M^2$ and $r$ by $M$ such that henceforth we will use the dimensionless parameters $b$ and $r$ throughout this paper.  
With \KR field as the source the above solution have been worked out previously in \cite{Kar:2002xa,SenGupta:2001cs}. For brevity we do not repeat the derivation here but simply mention the results.
From the form of the above metric it is clear that the presence of the Kalb-Ramond field does not lead to an exact black hole solution but results in perturbations over the Schwarzschild scenario by various powers of the axion parameter $b$.
The solution however is valid for all distances, viz, from the event horizon to infinity. This can be confirmed directly from the derivation of the above metric \cite{Kar:2002xa}.  
Since we are considering a spherically symmetric scenario the axion field $\Phi$ depends only on the radial coordinate $r$. Consequently, from \ref{5} it is clear that the only non-zero component of the \KR field strength tensor is $H^{023}$. The energy density corresponding to the Kalb-Ramond field is then given by $H^{023}H_{023}=h(r)^2$. By solving the gravitational field equations and the equations of motion for the \KR field  it can be shown that $h(r)$ assumes the form \cite{Kar:2002xa,SenGupta:2001cs},
\begin{align}
\label{6c}
h(r)=\sqrt{\frac{3b}{\kappa}}\frac{1}{r^2}\bigg[1+\frac{2}{r}+\frac{4}{r^2}-\bigg(\frac{8+b}{r^3}\bigg)+\bigg(\frac{16+6b}{r^4}\bigg) + ~...\bigg]
\end{align}
From \ref{6c} one can relate that the the parameter $b$ in \ref{6a} and \ref{6b} is associated with the energy density corresponding to the \KR field or the axion.
From \ref{6c} it is clear that at large distances, the \KR field energy density vanishes and we get back to the general relativistic scenario, which is also supported by the form of the solution of the metric (\ref{6a} and \ref{6b}). Therefore, this parametrisation of the metric and the \KR field energy density is valid for all distances, although its effect becomes prominent in the near horizon regime of the black hole.

It is important to note that in order for \ref{6} to represent a black hole solution, there must be an event horizon. The radius of the horizon $r_h$ is given by solving for $g^{rr}=e^{-\lambda(r)}=0$, which yields,
\begin{align}
r_{h}=1\pm \sqrt{1-3b}
\label{7}
\end{align}
if we truncate \ref{6b} upto the leading order term in $b$. The event horizon $r_{eh}$ is given by the positive root of \ref{7}.
 
In the next section we will consider the geodesic motion of the photons in the background given by \ref{6} which will enable us to derive the shape and size of the black hole shadow. It is important to note that the observation of the shadow directly probes the near horizon regime of black holes where $r$ is small. Therefore, although the leading order term with the axion appears as $1/r^3$ correction to the Schwarzschild scenario its impact on the observed shadow is expected to be significant. 

\section{Geodesic motion of photons and the shadow of a black hole}\label{S3}
The shadow of a black hole refers to the set of directions in the local sky from where electromagnetic radiation just escapes the black hole event horizon and reaches the observer on Earth \cite{Cunha:2018acu,Vries_1999,Gralla:2019xty,Abdujabbarov:2015xqa,Abdujabbarov:2016hnw}. When light from a distant astrophysical object or the accretion disk surrounding the black hole arrives in the vicinity of the event horizon, a part of it gets trapped inside the horizon while another part escapes to infinity. This results in a lack of radiation in the observer's sky leading to a dark patch in the image of the black hole, known as the black hole shadow. The outline of the shadow testifies the signatures of strong gravitational lensing of nearby radiation and hence the shape and size of the shadow can reveal valuable information regarding the nature of strong gravity near the black hole \cite{Gralla:2019xty,Bambi:2019tjh,Hioki:2009na,Vagnozzi:2019apd,Banerjee:2019nnj}. Consequently, the image of a black hole can be used as a potential probe to estimate the deviation from \gr. 

While the shape of the shadow bears imprints of the background geometry, the size of the shadow scales directly with its mass, reduces with increase in distance and also exhibits dependence on the background spacetime. For example a non-spinning black hole always casts a circular shadow \cite{Cunha:2018acu,Vries_1999}. In this case the size of the shadow can be used to investigate the deviation from the Schwarzschild scenario in \gr\ \cite{Cunha:2018acu,Vries_1999}.    
Introducing spin to black holes incurs deviation from the circular shape and this has been studied extensively in the past both in the context of \gr\ and alternative gravity models \cite{Cunha:2018gql,Gralla:2019xty,Bambi:2019tjh,Hioki:2009na,Vagnozzi:2019apd,Banerjee:2019nnj,Mizuno:2018lxz,Roy:2019esk}. However, it is important to note that the deviation from circularity becomes apparent only when the angle of inclination of the observer with respect to the rotation axis of the black hole becomes appreciable, i.e., an observer viewing a rotating black hole with zero inclination angle will always see a circular shadow \cite{Cunha:2018acu,Vries_1999}. 

In the next section we will derive the contour of the black hole shadow in the presence of a general static, spherically symmetric and asymptotically flat metric given by \ref{6} and subsequently we will consider the special case with axionic hairs where the metric components are given by \ref{6a} and \ref{6b}. 
%For a rotating black hole the shape of 

\subsection{Structure of black hole shadow in a general spherically symmetric metric} 
\label{S3a}
In this section we will work out the structure of the black hole shadow in a general static, spherically symmetric background given by \ref{6}. For this purpose we will study geodesic motion of photons in this spacetime. We consider a geodesic with an affine parameter 
$\lambda$ such that the tangent vector is $u^{\m}=\dot{x}^{\m}=dx^{\m}/{d\lambda}$. The Lagrangian $\mathcal{L}$ corresponding to the motion of test particles assumes the form,
\begin{align}\label{8}
  \mathcal{L}(x^{\m},\dot{x}^{\m})=\dfrac{1}{2}g_{\m\n}\dot{x}^{\m}\dot{x}^{\n}
  \end{align}
such that the action $S$ representing the motion of test particles satisfies the Hamilton-Jacobi equation given by,
\begin{align}
\label{9}
 \mathcal{H}(x^{\m},p_{\m})+\dfrac{\partial S}{\partial\lambda}=0
\end{align}
where,
 \begin{align}\label{10}
  \mathcal{H}=\dfrac{1}{2}\, g^{\m\n} p_{\m}p_{\n}=\dfrac{k}{2}
  \end{align}
is the Hamiltonian, $k$ is a constant representing the rest mass of the test particles 
(which is zero for photons) and $p_{\m}$ is the conjugate momentum corresponding to the coordinate $x^{\m}$ and is given by,
\begin{align}
\label{11}
  p_{\m}=\frac{\partial S}{\partial x^\mu}=\dfrac{\partial \mathcal{L}}{\partial \dot{x}^{\m}}=g_{\m\n}\dot{x}^{\m}
  \end{align}
Since the metric does not depend explicitly on $t$ and $\phi$, the energy $E$ and the angular momentum $L$ of the photons are conserved. These constants of motion are given by,  
 \begin{align}
 \label{12a}
  E=-g_{tt}u^{t}=-p_{t}~~~\rm and
  \tag{12a}
  \end{align}
 \begin{align}
  \label{12b}
  L=g_{\phi\phi}u^{\phi}=p_{\phi}
  \tag{12b}
  \end{align}
respectively. The action $S$ in \ref{11} can therefore be integrated with the help of \ref{12a} and \ref{12b} such that
\begin{align}
\label{13}
 S=-E t +L\phi + \bar{S}(r,\theta)
 \tag{13}
\end{align}
where in the case of a static, spherically symmetrically metric like \ref{6}, 
$\bar{S}(r,\theta)$ turns out to be separable in $r$ and $\theta$ with, $\bar{S}(r,\theta)=S^{r}(r)+S^{\theta}(\theta)$. We also note that with the help of \ref{11}, \ref{10} can be written as,
\begin{align}
\label{14}
-e^{-\nu(r)}r^2E^2 + e^{-\lambda(r)}r^2\bigg(\frac{d{S^r}}{dr}\bigg)^2 + L^2= -\bigg(\frac{d{S^\theta}}{d\theta}\bigg)^2 - L^2\rm{cot}^2(\theta)=-C
\tag{14}
\end{align}
where the separation constant $C$, known as the Carter constant represents a third constant of motion \cite{Carter:1968rr}. Therefore the geodesic equations for $r$ and $\theta$ are given by,
\begin{align}
\label{15}
\left(\dfrac{dS^{r}}{dr}\right)^{}&=\sqrt{e^{\lambda(r)}\left(-\dfrac{C}{r^2}-\frac{L^2}{r^2}+e^{-\n(r)} E^2\right)}= E\, \sqrt{V(r) }=g_{rr}\dot{r}~~~~~\rm{and}
\tag{15}
\end{align}
\begin{align}
\label{16}
  \left(\dfrac{dS^{\theta}}{d\theta}\right)&=\sqrt{C-L^2{\cot^2 \theta}} = E\, \sqrt{\Theta (\theta)}=g_{\theta\theta}\dot{\theta}
\tag{16}
\end{align}
respectively, where
\begin{align}
\label{17}
V(r)=-\frac{e^{\lambda(r)}\chi}{r^2}-\frac{e^{\lambda(r)}l^2}{r^2}+e^{\lambda(r)-\nu(r)} 
\tag{17}
\end{align}
represents the effective potential in which the photon moves, while
\begin{align}
\label{18}
{\Theta}(\theta)=\chi-l^2 \cot^2\theta
\tag{18}
\end{align}
such that $\chi=C/E^2$ and $l=L/E$. 
The radius of the photon sphere $r_{ph}$ corresponds to the condition where $\dot{r}$ vanishes and the effective potential $V(r)$ has an extrema. This is generally a maxima, which corresponds to an unstable equilibrium of the photon. Given a slight perturbation the photon either falls into the horizon or escapes to infinity. Due to this reason the photon sphere plays a crucial role in determining the boundary of the black hole shadow. 

Therefore, $r_{ph}$ is obtained by solving $V(r)=V'(r)=0$, such that the above conditions yield
\begin{align}
\label{19}
\chi+l^2=r_{ph}^2 e^{-\nu(r_{ph})}~~~~~\rm{and}
\tag{19}
\end{align}
\begin{align}
\label{20}
\chi+l^2=\frac{1}{2}r_{ph}^3 e^{-\nu(r_{ph})}\nu^\prime
\tag{20}
\end{align}
respectively. The photon sphere in an arbitrary spherically symmetric metric is therefore obtained by solving for $r$ in the following equation,
\begin{align}
\label{21}
r_{ph} \,\n'(r_{ph})=2
\tag{21}
\end{align}
In order to derive the contour of the black hole shadow in the observer's sky one considers the projection of the photon sphere in the image plane \cite{Bardeen:1973tla}. Note that the largest positive radius obtained by solving \ref{21} is relevant for the computation of the shadow outline \cite{Vries_1999,Cunha:2018acu}. The locus of the shadow boundary is denoted in terms of two celestial coordinates $\alpha$ and $\beta$ which are related to $l$ and $\chi$ \cite{Bardeen:1973tla,Vries_1999}. 
This can be understood by expressing the metric in terms of the tetrads which for a spherically symmetric background assumes the form,

\begin{align}
\label{22}
\tag{22}
g_{\mu\nu}=e^{(a)}_\mu e^{(b)}_\nu \eta_{ab}~~~~~\rm{where}
\end{align}
%\begin{subequations}
\begin{alignat}{4}
\label{23}
e^{(t)}_{\m}&=(e^{\n/2}, 0, 0, 0) \tag{23a}\\
e^{(r)}_{\m}&=(0,e^{\lambda/2},0,0)  \tag{23b}\\
e^{(\theta)}_{\m}&=(0,0,r,0)   \tag{23c}\\
e^{(\phi)}_{\m}&=(0,0,0,r \sin\theta)  \tag{23d}
\end{alignat}
%\end{subequations}
Particularly, the apparent velocity $v_{(\theta)}$ of the photon in the $\theta$ direction and $v_{(\phi)}$ of the photon in the $\phi$ direction in the local rest frame are given by,

%\begin{subequations}
\begin{align}
\label{24}
v_{_{(\theta)}}&=\dfrac{u_{\m}e^{\m}_{(\theta)}}{u_{\m}e^{\m}_{(t)}}=\dfrac{p_{\theta}e^{\theta}_{(\theta)}}{p_{t}e^{t}_{(t)}}=\dfrac{\mp \sqrt{\Theta(\theta)}e^{\n/2}}{r}~~~~~\rm{and}
\tag{24}
\end{align}

\begin{align}
\label{25}
v_{_{(\phi)}}&=\dfrac{u_{\m}e^{\m}_{(\phi)}}{u_{\m}e^{\m}_{(t)}}=\dfrac{p_{\phi}e^{\phi}_{(\phi)}}{p_{t}e^{t}_{(t)}}=-\dfrac{le^{\n/2}}{r\sin\theta}
\tag{25}
\end{align}
%\end{subequations}
respectively. 
An observer located at a distance $r_0$ with an inclination angle $\theta_0$ will perceive that the celestial coordinates are given by,
\begin{align}
\label{26}
\beta=\lim_{r_0\to\infty} r_0 v_{_{(\theta)}}(r_0,\theta_0)=\mp \sqrt{\Theta(\theta_0)} 
\tag{26}
\end{align}
\begin{align}
\label{27}
\alpha=\lim_{r_0\to\infty} r_0 v_{_{(\phi)}}(r_0,\theta_0)=-\frac{l}{\sin\theta_0}
\tag{27}
\end{align}
Note that $r_0$ do not appear in the expression for $\alpha$ and $\beta$ since the metric is assumed to be asymptotically flat. Using \ref{18} it can be shown that 
\begin{align}
\label{28}
\alpha^2 + \beta^2=\chi+l^2={r_{sh}^2}
\tag{28}
\end{align} 
which shows that the shadow is circular in shape where the dependence of its radius $r_{sh}$ on $r_{ph}$ is given by \ref{19}.
The above discussion clearly elucidates that for any general static, spherically symmetric and asymptotically flat metric the shadow is circular in shape. Further, the radius of the shadow depends only on the $g_{tt}$ component of the metric and is independent of the distance $r_0$ and the inclination angle $\theta_0$ of the observer.

\subsection{Structure of black hole shadow in presence of Kalb-Ramond field}
\label{S3b}
In this section we will compute the contour of the black hole shadow by considering specifically the spherically symmetric metric with axionic hairs (\ref{6} with \ref{6a} and \ref{6b} as metric elements).
As discussed in the previous section, the radius of the photon sphere is obtained by solving \ref{21} which for our specific case leads to
\begin{align}
\label{29}
2 r^3-6 r^2+5b=0
\tag{29}
\end{align}
Depending on the value of the axion parameter $b$, \ref{29} can have three distict real roots, one distinct and one coincident real root or a single distinct real root. 
The conditions for the above are enlisted below:
\begin{align*}
\begin{cases}
   \text{Three distinct real roots} & 0<b<1.6   \\
\text{Only one real root} & b<0;\,\,b>1.6\\
\text{One real and one coincident root} & b=0;\,\,b=1.6
\end{cases}
\end{align*}
The roots of this equation can be obtained analytically by using Cardano's method \cite{nickalls_2006}. 

\begin{figure}[h]%
    \centering
    \subfloat[Radius of photon sphere vs. b]{{\includegraphics[width=7.5cm]{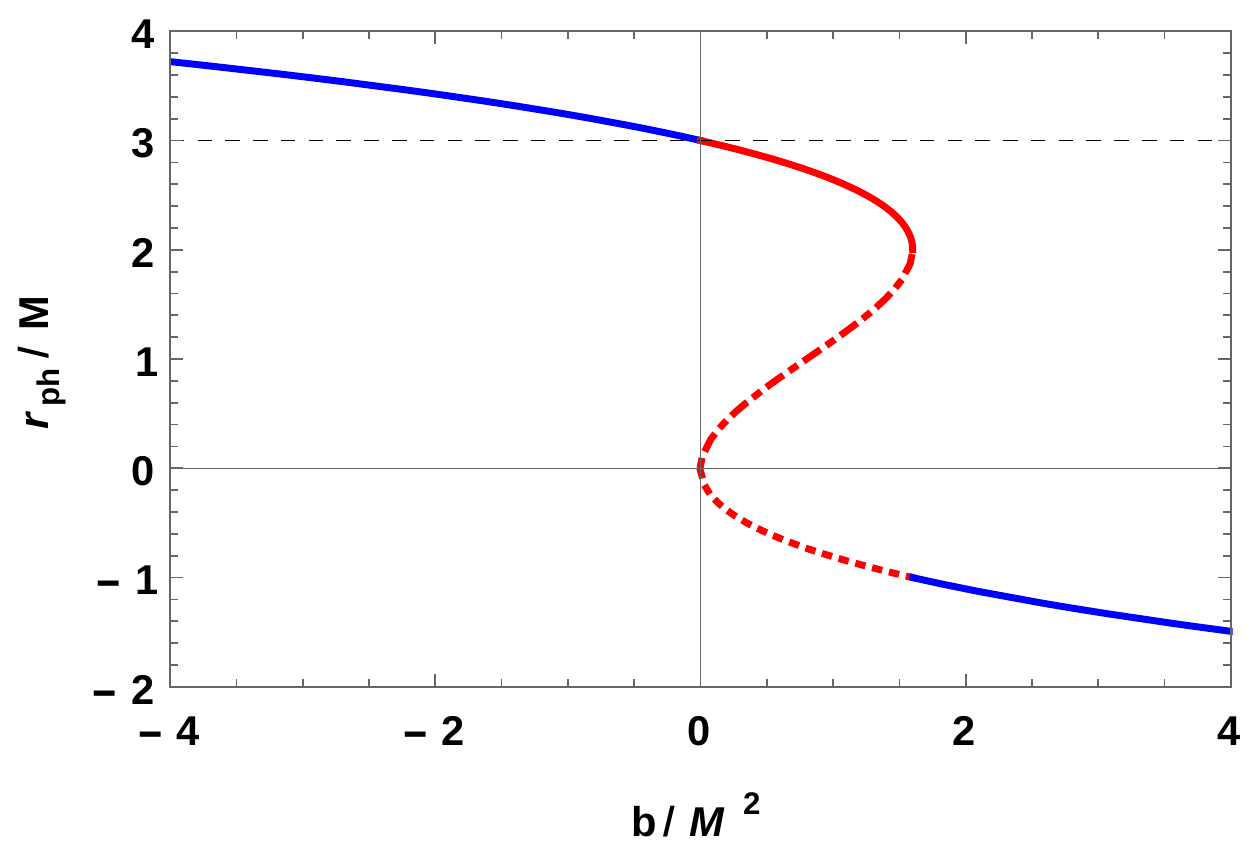} } \label{P2Fig1a}}
    \qquad
    \subfloat[Radius of shadow vs. b]{{\includegraphics[width=7.4cm]{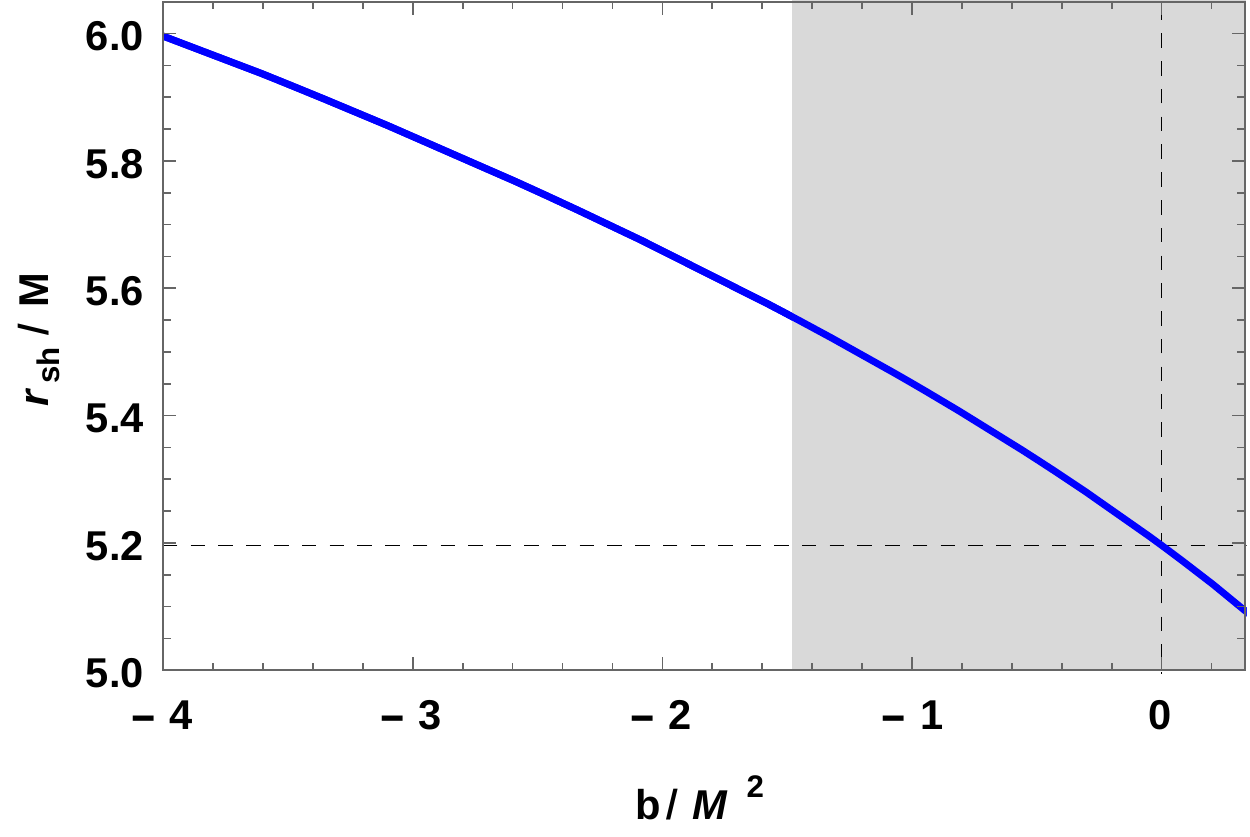} } \label{P2Fig1b}}
    \caption{Dependence of radius of photon sphere and the radius of shadow with the axion parameter }
    \label{P2Fig1}%
\end{figure}

When $0<b<1.6$, the three real roots are given by,
\begin{align}
\label{30a}
r_1=1+2\cos\bigg[\frac{1}{3}\cos^{-1}B\bigg]
\tag{30a}
\end{align}
\begin{align}
\label{30b}
r_2=1-\cos\bigg[\frac{1}{3}\cos^{-1}B\bigg]+\sqrt{3}\sin\bigg[\frac{1}{3}\cos^{-1}B\bigg] 
\tag{30b}
\end{align}
\begin{align}
\label{30c}
r_3=1-\cos\bigg[\frac{1}{3}\cos^{-1}B\bigg] -\sqrt{3}\sin\bigg[\frac{1}{3}\cos^{-1}B\bigg] 
\tag{30c}
\end{align}
with $|B|<1$ where $B=-1+5b/4$. 
In the event, $b=0$ or $b=1.6$, $r_2$ coincides with $r_3$, while $r_1$ corresponds to another distinct real root. 
When $b<0$ or $b>1.6$, which is identical to the situation with $|B|>1$, there is only one real root which is given by,
\begin{align}
\label{31}
r_0=1+\bigg[\bigg\lbrace |B|+ \sqrt{B^2-1}\bigg\rbrace^{1/3}+\bigg\lbrace |B|+ \sqrt{B^2-1}\bigg\rbrace^{-1/3}\bigg]
\tag{31}
\end{align} 
The radius of the photon sphere is depicted clearly in \ref{P2Fig1a}. 
 %%%%%%%%%%%%%%%%%
%%%%%%%%%%%%%%%%%%%
In the figure, the blue solid line corresponds to the condition $|B|>1$ while the red curves constitutes the situation with $|B|<1$. Since the latter consists of three distinct radii, $r_1$, $r_2$ and $r_3$ are marked with solid, dot-dashed and dotted red lines respectively. Among these the greatest positive root is taken for the computation of the shadow radius. It is important to note that for $b>1.6$ only one real root exists but the root is always negative (which is unphysical since the photon sphere cannot have a negative radius) and hence we conclude that $b$ can never exceed this limit. The maximum value of $b$ is further reduced from the consideration that for the metric to represent a black hole the event horizon has to be real, which from \ref{7} implies $b<1/3=b_{max}$. Henceforth, we will consider $b_{max}$ to be the upper limit of $b$.

For the region $b<0$, \ref{29} has only one positive real root $r_0$ which increases as $b$ decreases. It is important to note that $b$ cannot assume arbitrarily large negative values since when $b\lesssim-1.48=b_{min}$ the radius of the event horizon $r_{eh}$ exceeds the photon sphere $r_{ph}$. Consequently, when $b$ is lower than $b_{min}$ photon circular orbits do not exist. This is an interesting feature the spacetime inherits due to the presence of the Kalb-Ramond background. 
Therefore, in the remaining discussion we will limit ourselves in the range $-1.48\lesssim b\lesssim 1/3$. Once the dependence of the photon sphere on $b$ is understood, we compute the radius of the shadow $r_{sh}$ in terms of the axionic parameter $b$ using \ref{19} and \ref{28}.  
The variation of $r_{sh}$ with $b$ is plotted in \ref{P2Fig1b} where we have shaded the theoretically allowed region of $b$. We also note that the shadow expands with negative values of $b$. This result will have interesting consequences from the observed shadow of M87* which we discuss in the next section.

We emphasize once again that since we are probing the near horizon regime, even the leading order correction to the $g_{tt}$ component of the Schwarzschild metric given by $b/r^3$, is expected to be significant near the photon sphere. This is in accordance with our analysis which reveals that the axion parameter plays a crucial role in affecting the radius of the black hole shadow compared to the Schwarzschild scenario (\ref{P2Fig1b}). 
Since we are interested in the small $r$ domain it might appear that one should also consider the higher order corrections to the metric (terms over and above the leading order corrections in the $g_{tt}$ and the $g_{rr}$ components) in \ref{6a} and \ref{6b} to study the impact of axion on black hole shadow and the radius of the event horizon.
However inclusion of these terms does not significantly affect our results since $b$ has a theoretical bound of $-1.48\lesssim b \lesssim1/3$ and in this domain $b/r <1$ (\ref{P2Fig1a}) which implies $b/{r^3}$ or $b/{r^2}$ are even less. This enables us to truncate the metric in \ref{6a} and \ref{6b} upto the leading order term. However, we verify this approximation explicitly by considering terms upto $1/r^4$ in both $g_{tt}$ and the $g_{rr}$ components of the metric and find that this has negligible effect on our results.

%Further note that for the metric to represent a black hole the event horizon has to be real which from \ref{7} implies $b<1/3=b_{max}$. Henceforth, we will consider $b_{max}$ to be the upper limit of $b$. 

%%%%%%%%%%%%%%%%%%%%%%%%%
%%%%%%%%%%%%%%%%%%%%%%%%%%
%%%%%%%%%%%%%%%%%%%%%%%%%

%%%%%%%%%%%%%%%%%%%%%%%%%%%%%%%%
%%%%%%%%%%%%%%%%%%%%%%%%%%%%%%%%
\section{Observed shadow of M87* and implications on axionic hair}
\label{S4}
Using the techniques of VLBI (Very Large Baseline Interferometry), the Event Horizon Telescope (EHT) Collaboration has recently released the image of the supermassive black hole M87* at the centre of the galaxy M87, thereby opening a new window to test gravity in the strong field regime \cite{Akiyama:2019cqa,Akiyama:2019brx,Akiyama:2019sww,Akiyama:2019bqs,Akiyama:2019fyp,Akiyama:2019eap}. Their analysis reveals that the angular diameter of the shadow of M87* is $(42\pm 3) \mu as$ exibiting a deviation from circularity $\Delta C<10\%$ and the axis ratio $\Delta A<4/3$ \cite{Akiyama:2019cqa}. This implies that the observed shadow is nearly circular which is further supported from the fact that the jet axis makes an angle of $17^\circ$ to the line of sight, which is taken to be the inclination angle of the the black hole \cite{Akiyama:2019cqa,Akiyama:2019fyp,Akiyama:2019eap}. We have already mentioned in \ref{S3} that non-circular shadows are only possible if a black hole is observed at high inclination angle. This therefore, justifies our choice of considering the spherically symmetric metric given by \ref{6}, as a first approximation. 
Hence, the only relevant observable in our context is the angular diameter of the shadow, $\Delta A$ and $\Delta C$ being trivially equal to one and zero respectively, satisfying the observed constraint.
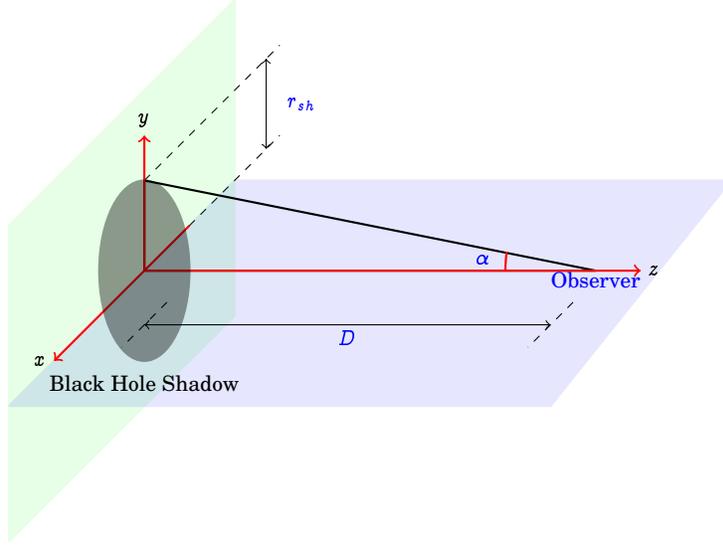
\begin{figure}[h]
\centering
\begin{tikzpicture}[thick, scale=0.6,every node/.style={scale=0.8}]
%\draw [red,shade](0,0) circle (.4cm);
%\draw[fill] (0,0) circle (.2cm);
\draw[thin,dashed] (0,0)--(3,3);
\draw[thin,dashed] (0,2)--(3,5);
\draw[fill,green,opacity=.1] (-3,-6)--(-3,1)--(2,6)--(2,-1)--(-3,-6);
\draw[fill,blue,opacity=.1] (-3,-3)--(2,2)--(13,2)--(9,-3)--(-3,-3);
%\draw[dashed,green] (-.9,-.9)--(10,0);
\draw[] (0,0)--(10,0);
\draw[] (0,2)--(10,0);
\draw[] (0,0)--(0,2);
\draw[thick,red,<-] (-2,-2)--(1,1);
\draw[thick,red,->] (0,0)--(0,3);
\draw[thick,red,->] (0,0)--(11,0);
\draw[thin,<->] (2.7,2.7)--(2.7,4.7);
\draw[thin,<->] (0,-1.2)--(9,-1.2);
\draw[thin,dashed] (0.5,-.7)--(-0.5,-1.7);
\draw[thin,dashed] (9.5,-.7)--(8.5,-1.7);
\node at (-2,-2) [left] {$x$};
\node at (0,3) [above] {$y$};
\node at (11,0) [right] {$z$};
\node at (3,3.7)[right] {{\color{blue}{$r_{sh}$}}};
\node at (4.5,-1.5)[] {{\color{blue}{$D$}}};
\node at (0,-2.5)[] {{\color{black}{\text{Black Hole Shadow}}}};
\node at (7.5,.25)[] {{\color{blue}{$\alpha$}}};
\node at (10,-0.2)[] {{\color{blue}{\text{Observer}}}};
\draw[fill,black,opacity=.4] ellipse ( 1 cm and 2cm);
\draw[red] (8,0) arc (180:172:3);
\end{tikzpicture}
\caption{Black hole shadow with angular diameter $2\alpha$}
\label{Angular_Radius}
\end{figure}

The angular diameter of the shadow depends not only on the background metric but also on the mass $M$ of the black hole and its distance $D$ from the observer. 
This has been illustrated in \ref{Angular_Radius} which shows that 
\begin{flalign}
\label{AD}
\tan \alpha \approx \alpha = \dfrac{r_{sh}}{D} \tag{32}
\end{flalign} 
where $2\alpha$ is the angular diameter. Since the distance between the black hole and the observer is much much greater than the radius of the shadow ($r_{sh}$), $\alpha$ is very small which justifies the approximation in \ref{AD}.
We have already expressed the radius of the shadow in terms of the metric parameter $b$ in \ref{6} (e.g. \ref{P2Fig1b}). 
Since the radius is in units of $GM/c^2$ the angular diameter scales directly with the black hole mass. Further, black holes at larger distances will cast smaller shadows.

In the previous section we have computed the dependence of the axion parameter $b$ on the radius of the photon sphere and the shadow. Therefore from the magnitude of the observed angular diameter we can comment on the observationally favored values of $b$.  We however, require independent measurements of the mass and distance of M87*. Based on stellar dynamics and gas dynamics measurements the mass of M87* is reported to be 
$M\sim 6.2^{+1.1}_{-0.5}\times 10^9 M_\odot$ \cite{Gebhardt:2011yw} and $M\sim 3.5^{+0.9}_{-0.3}\times 10^9 M_\odot$ \cite{Walsh:2013uua} respectively while the distance of the source is reported to be $D=(16.8\pm0.8)~\textrm{Mpc}$ \cite{Blakeslee:2009tc,Bird:2010rd,Cantiello:2018ffy} from stellar population measurements. Moreover, the mass of the object reported by the EHT Collaboration derived from the angular diameter of the shadow assuming \gr\ is $M=(6.5\pm 0.7)\times 10^{9}~M_{\odot}$ \cite{Akiyama:2019cqa,Akiyama:2019fyp,Akiyama:2019eap} and hence should not be used to constrain the value of $b$ or other alternate gravity models. Further, the observed emission ring is actually expected to be $\sim 10\%$ larger than the true shadow size which is supported by multiple simulations of the accretion flow around M87* \cite{Akiyama:2019eap}.

\begin{figure}[h!]%
    \centering
    \hspace{-1.5cm}
    \subfloat[Angular diameter versus b with observed values $42\pm 3 \mu as$ marked in blue]{{\includegraphics[width=8.1cm]{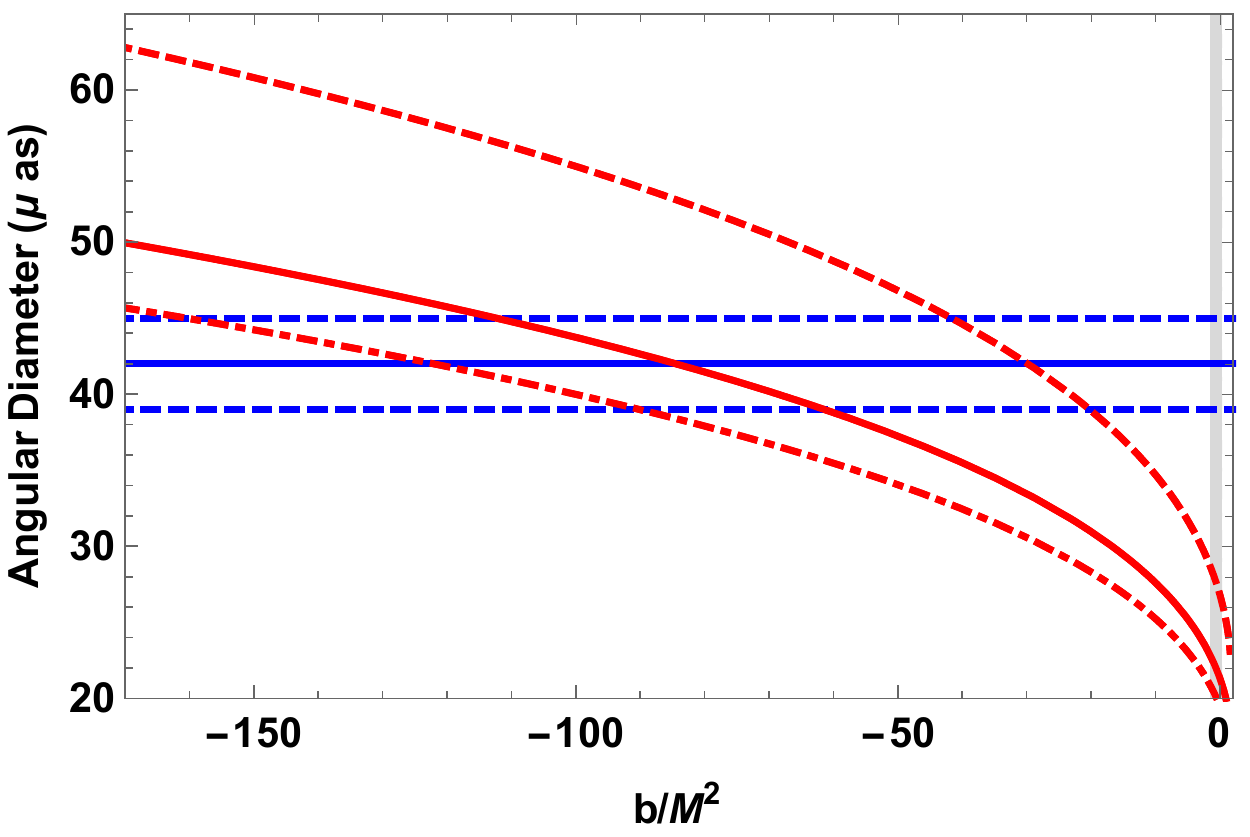} } \label{P2Fig3a}}
    \qquad
    \subfloat[Angular diameter versus b with observed values (with 10\% offset, $37.8\pm 2.7 \mu as$) marked in blue]{{\includegraphics[width=8.1cm]{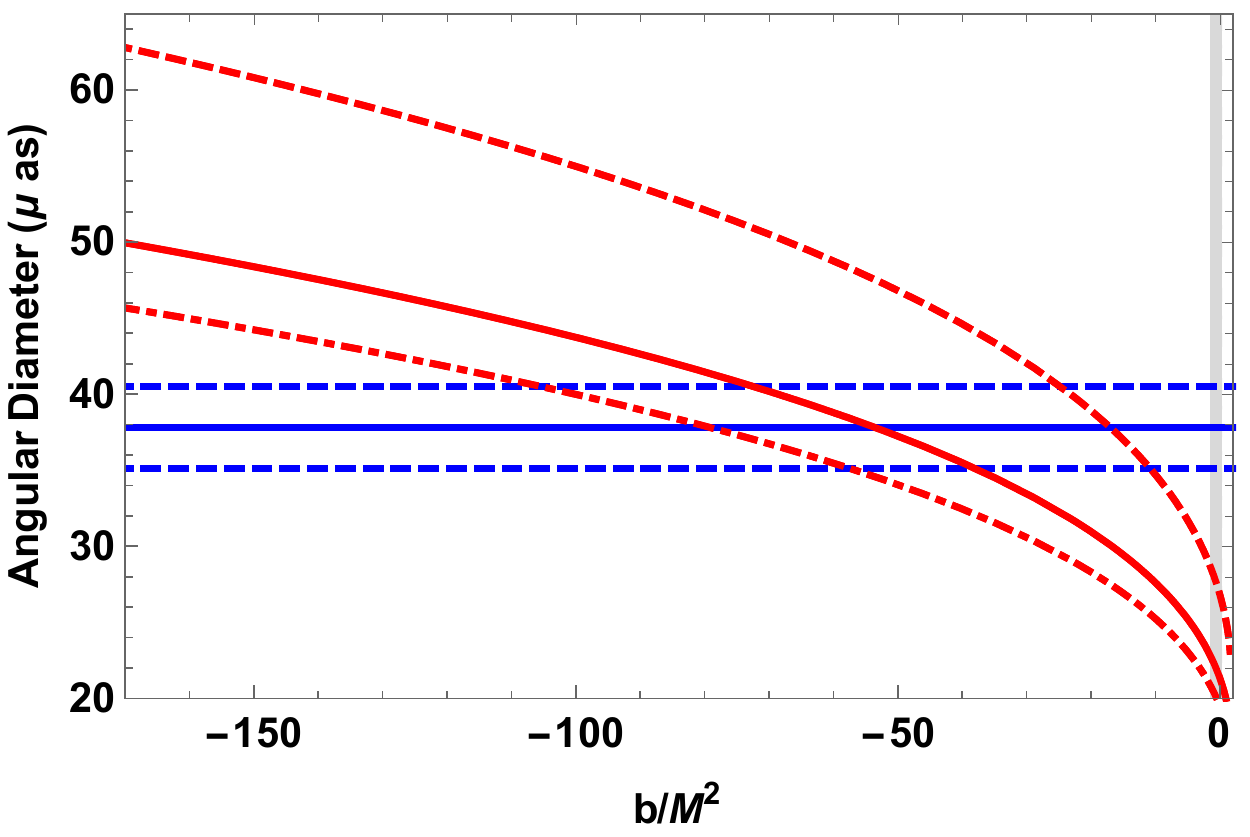} } \label{P2Fig3b}}
    \caption{Variation of the angular diameter of M87* (\ref{AD}) with axion parameter $b$ assuming $M=3.5^{+1.1}_{-0.5}\times 10^{9}M_\odot$ and $D=16.8~\textrm{Mpc}$}
    \label{P2Fig3}%
\end{figure}

\begin{figure}[h!]%
    \centering
    \hspace{-1.5cm}
    \subfloat[Angular diameter versus b with observed values $42\pm 3 \mu as$ marked in blue]{{\includegraphics[width=8.1cm]{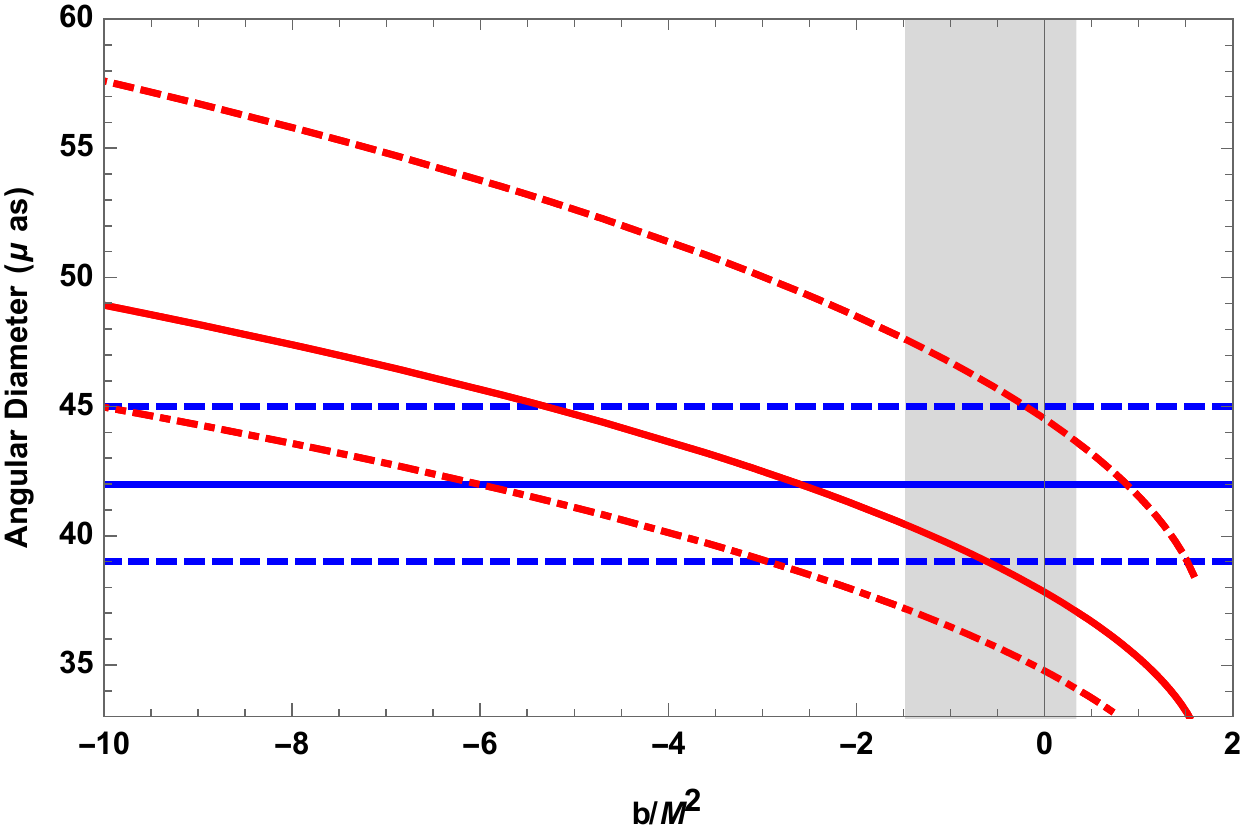} } \label{P2Fig4a}}
    \qquad
    \subfloat[Angular diameter versus b with observed values (with 10\% offset, $37.8\pm 2.7 \mu as$) marked in blue]{{\includegraphics[width=8.1cm]{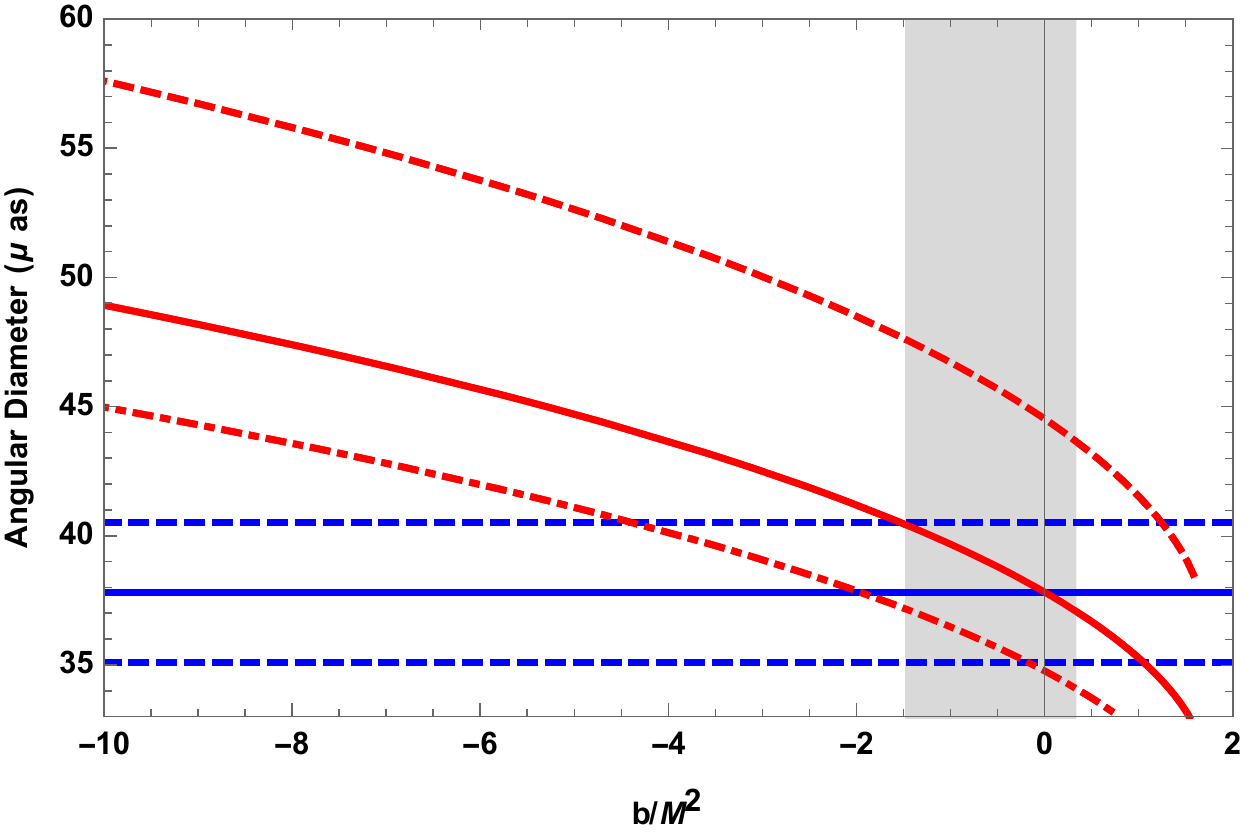} } \label{P2Fig4b}}
    \caption{Variation of the angular diameter of M87* (\ref{AD}) with axion parameter $b$ assuming $M=6.2^{+1.1}_{-0.5}\times 10^{9}M_\odot$ and $D=16.8~\textrm{Mpc}$}
    \label{P2Fig4}%
\end{figure}

\begin{figure}[h!]%
    \centering
   \hspace{-1.5cm} 
    \subfloat[Angular diameter versus b with observed values $42\pm 3 \mu as$ marked in blue]{{\includegraphics[width=8.1cm]{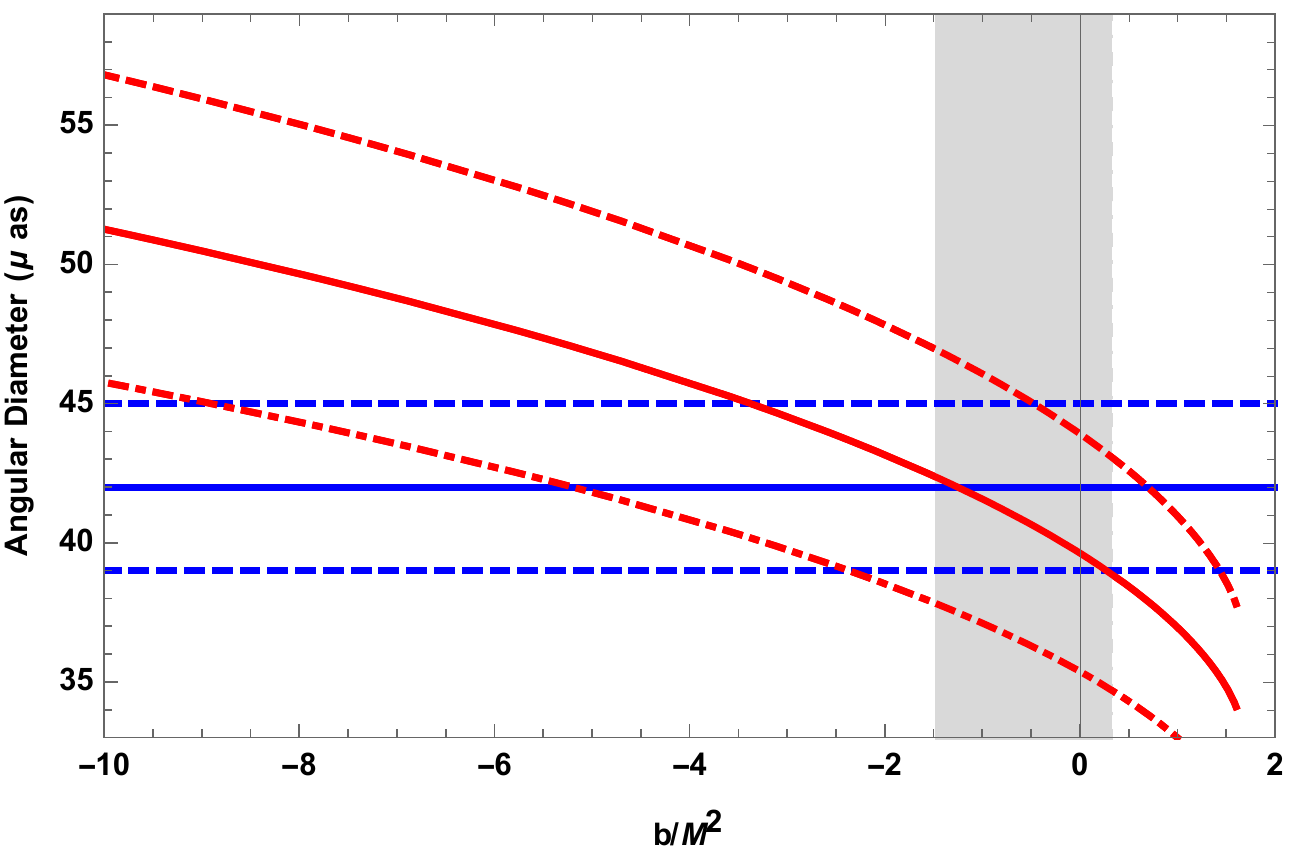} } \label{P2Fig5a}}
    \qquad
    \subfloat[Angular diameter versus b with observed values (with 10\% offset, $37.8\pm 2.7 \mu as$) marked in blue]{{\includegraphics[width=8.1cm]{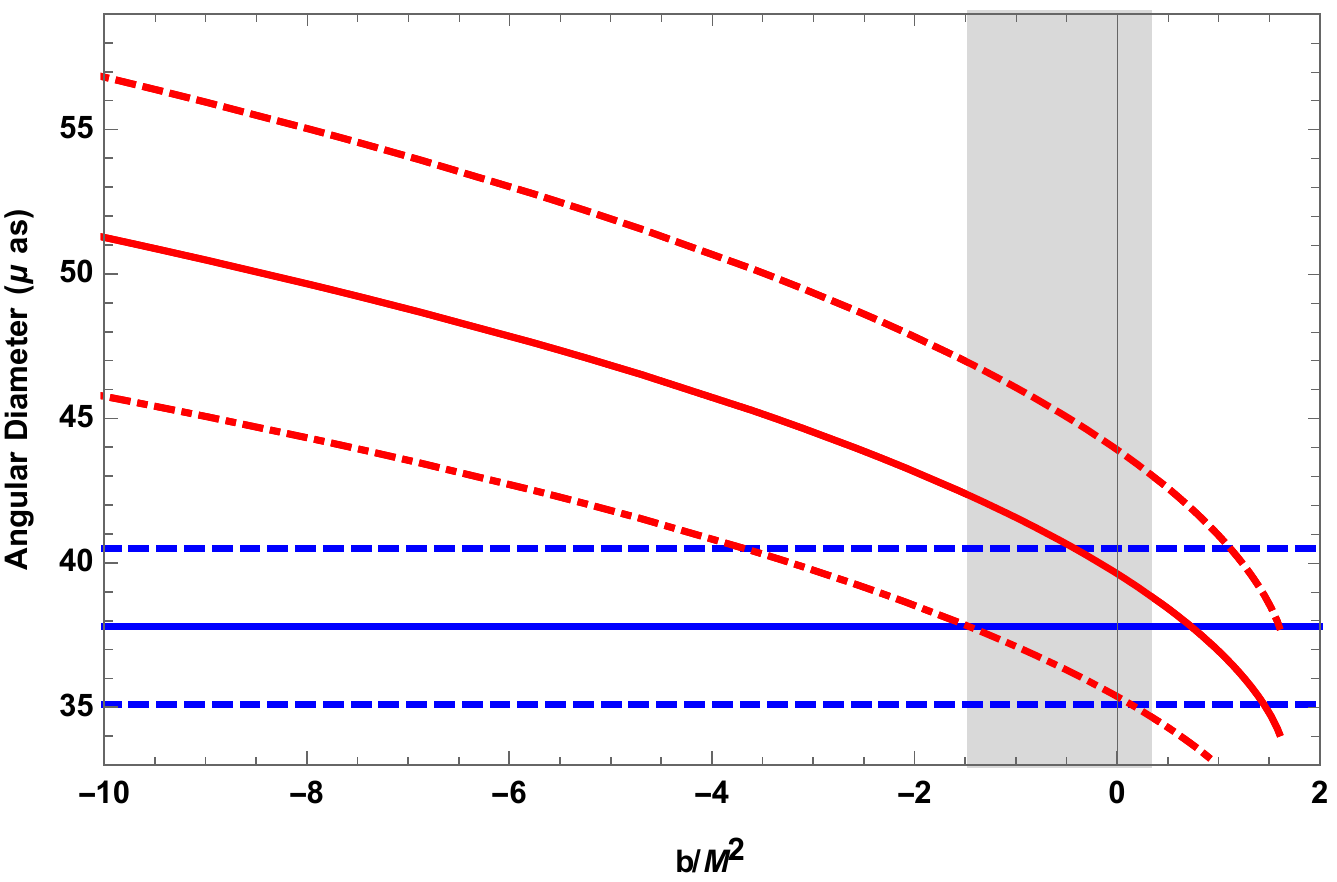} } \label{P2Fig5b}}
    \caption{Variation of the angular diameter of M87* (\ref{AD}) with axion parameter $b$ assuming $M=6.5\pm0.7\times 10^{9}M_\odot$ and $D=16.8~\textrm{Mpc}$}
    \label{P2Fig5}%
\end{figure}

Using these masses and distance, the theoretical angular diameter of M87* \ref{AD} is plotted with $b$ assuming the mass estimated from gas dynamics observations ($M=3.5^{+0.9}_{-0.3}\times 10^{9}M_\odot$) in \ref{P2Fig3} and from stellar dynamics measurements ($M=6.2^{+1.1}_{-0.5}\times 10^{9}M_\odot$) in \ref{P2Fig4}. In \ref{P2Fig5} the variation of the angular diameter with $b$ is plotted considering $M=6.5\pm0.7\times 10^{9}M_\odot$, which is the mass of M87* deduced from the angular diameter of the observed shadow assuming \gr. Since this is not an independent mass estimation it should not be used for constraining the value of $b$ from observations related to the shadow, although it can serve the purpose of comparison with the two independent mass measurements. In \ref{P2Fig3a}, \ref{P2Fig4a} and \ref{P2Fig5a} the observed angular diameter of $42\mu as$ is marked with solid blue line while the error of $\pm 3\mu as$ about the centroid value are depicted with blue dashed lines. Figures \ref{P2Fig3b},  \ref{P2Fig4b} and \ref{P2Fig5b} are the same as \ref{P2Fig3a}, \ref{P2Fig4a} and \ref{P2Fig5a} respectively, except that the observed angular diameter with 10\% offset i.e. $37.8\pm 2.7 \mu as$ are marked with blue lines. In each of the six figures, the theoretical angular diameters are plotted assuming the error bars in the masses such that
the solid red line corresponds to the centroid value, the dashed red line (above the solid line) is associated with the positive error bar and the dot-dashed red line (below the solid line) corresponds to the negative error bar in the mass.
It is important to note that the angular diameter is inversely proportional to the distance of the source (\ref{AD}), which turns out to be $16.8\pm0.8 $Mpc for M87* (estimated based on stellar population measurements). All the figures (\ref{P2Fig3}, \ref{P2Fig4} and \ref{P2Fig5}) mentioned above are plotted with the centroid value of the distance, i.e. $16.8  $Mpc. Also, the theoretically allowed values of $b$ ($-1.48\lesssim b \lesssim1/3$) are shaded in all the six figures.

\vskip 1mm
\noindent
%\rule{0.95\textwidth}{2pt}\\
\begin{table}[]
\caption{Values of axion parameter $b$ required to explain the observed angular diameter of $42\pm 3\mu as$ and $37.8\pm 2.7 \mu as$ (highlighted in blue). The latter consists of the deviation in the observed angular diameter when a maximum offset of $10\%$ is allowed. The masses used for computing the theoretical angular diameter are reported as Serial No. 1, 2 and 3. For $M=3.5^{+0.9}_{-0.3}\times 10^{9}M_\odot$ the $b$ values are obtained from \ref{P2Fig3} when the theoretical angular diameter equals the observed angular diameter of $42\pm 3\mu as$ (\ref{P2Fig3a}) and $37.8\pm 2.7 \mu as$ (\ref{P2Fig3b}). Similarly, for $M=6.2^{+1.1}_{-0.5}\times 10^{9}M_\odot$, the $b$ values are obtained from \ref{P2Fig4a} and \ref{P2Fig4b} while for $M=6.5\pm0.7\times 10^{9}M_\odot$ the $b$ values correspond to \ref{P2Fig5a} and \ref{P2Fig5b}}.

\vskip 5mm
\begin{tabular}{|c|c|c|c|c|c|c|c|}
\hline
\multirow{3}{*}{Serial No.} & \multirow{3}{*}{\begin{tabular}[c]{@{}c@{}}Mass\\ (In units of $10^{9}M_{\odot}$)\end{tabular}} & \multicolumn{6}{c|}{Value of parameter $b$ for given values of angular diameter}    \\ \cline{3-8} 
                            &                                                                                                           & \multicolumn{3}{c|}{Set 1} & \multicolumn{3}{c|}{Set 2} \\ \cline{3-8} 
                            &                                                                                                           & {\cellcolor[HTML]{96FFFB}\color{blue}45 $\mu as$}      & {\cellcolor[HTML]{96FFFB}\color{blue}42 $\mu as$}      & {\cellcolor[HTML]{96FFFB}\color{blue}39 $\mu as$}     & {\cellcolor[HTML]{96FFFB}\color{blue}40.5 $\mu as$}     & {\cellcolor[HTML]{96FFFB}\color{blue}37.8 $\mu as$}    &{\cellcolor[HTML]{96FFFB}\color{blue}35.1 $\mu as$}    \\ \hline

\multirow{3}{*}{1}          & 3.5+0.9                                                                                                   & -42.8   & -30.8   & -21    & -25.2   & -17.4   & -11.2  \\ \cline{2-8} 
                            & 3.5                                                                                                       & -115    & -86.2   & -63    & -73     & -54     & -38.4  \\ \cline{2-8} 
                            & 3.5-0.3                                                                                                   & -162.4  & -124.2  & -91.4  & -107    & -81     & -58.8  \\ \hline
                            
\multirow{3}{*}{2}          & 6.2+1.1                                                                                                   & -0.2    & 0.85    & 1.47   & 1.25    & 1.6     &        \\ \cline{2-8} 
                            & 6.2                                                                                                       & -5.4    & -2.65   & -0.55  & -1.55   & 0       & 1.05   \\ \cline{2-8} 
                            & 6.2-0.5                                                                                                   & -10     & -6.2    & -2.95  & -4.5    & -2      & -0.25  \\ \hline                            
                            
 \multirow{3}{*}{3}          & 6.5+0.7                                                                                                   & -0.55   & 0.7     & 1.4    & 1.1     & 1.6     &        \\ \cline{2-8} 
                            & 6.5                                                                                                       & -3.45   & -1.3    & 0.25   & -0.45   & 0.7     & 1.4    \\ \cline{2-8} 
                            & 6.5-0.7                                                                                                   & -8.8    & -5.3    & -2.4   & -3.85   & -1.45   & 0.15   \\ \hline

\end{tabular}
\label{T1}
\end{table}

Depending on the mass used, the value of $b$ required to explain the observed angular diameter varies. \ref{T1} enlists those values of $b$ where the observed angular diameter of $42\pm 3 \mu as$ (Set 1) and $37.8\pm 2.7 \mu as$ (Set 2) (highlighted in blue in \ref{T1}) are reproduced, assuming the three mass estimations of M87* (denoted by Serial No. 1, 2 and 3 in \ref{T1}). These values of $b$ are essentially obtained at the points of intersection of the blue lines and the red curves in \ref{P2Fig3}, \ref{P2Fig4} and \ref{P2Fig5}. 

From \ref{T1} we note that, if $M=3.5^{+0.9}_{-0.3}\times 10^{9}M_\odot$ is assumed, only negative values of $b$ can explain the observed angular diameter even when the $10\%$ offset in the angular diameter is considered. This is because the angular diameter scales directly with the mass and the shadow radius $r_{sh}$ (\ref{AD}) which increases for a negative $b$ compared to the general relativistic scenario ($b=0$) (\ref{P2Fig1b}). Therefore, when a smaller mass of the source is considered a more negative value of $b$ is required to address the observed angular diameter. For the same reason, when larger masses i.e., $M=6.2^{+1.1}_{-0.5}\times 10^{9}M_\odot$ or $M=6.5\pm0.7\times 10^{9}M_\odot$ are considered, the observationally favored values of $b$ are less negative compared to the case with $M=3.5^{+0.9}_{-0.3}\times 10^{9}M_\odot$. Further, when the maximum  offset of $10\%$ in the angular diameter (i.e., $37.8\pm 2.7 \mu as$) is considered or the positive error bar in the masses are considered, $b=0$ comes within the error bars. It is evident from \ref{T1} and \ref{P2Fig4b} and \ref{P2Fig5b} that corresponding to observed angular diameter of $35.1\mu as$, no value of $b$ can explain the data when the positive errors associated with masses $M=6.2 \times 10^{9} M_\odot$ and $M=6.5\times 10^{9}M_\odot$ are considered. This is because, with these masses the value of $b$ required to explain the observed angular diameter of $35.1\mu as$ is $b>1.6$ which is physically prohibited since the photon sphere assumes a negative radius beyond this value (\ref{S3b}). These entries in the table are therefore left blank. 

While it is apparent from \ref{T1} that a negative $b$ explains the observed angular diameter better, a chi-squared analysis taking into account the uncertainties in the mass, distance and the accretion flow model is performed to strengthen our conclusion. The chi-square is given by,

\begin{align}
\label{33}
\chi^2(b)=\sum_{i}\frac{\left \lbrace\mathcal{O}-\mathcal{T}_i(\left\lbrace M_k \right\rbrace,\left\lbrace D_k \right\rbrace ,b)\right\rbrace^2}{\sigma^2}
\tag{33}
\end{align}
where $\mathcal{O}$ corresponds to the observed angular diameter of $42\mu as$ with a standard deviation $\sigma=\pm 3 \mu as$, while $\mathcal{T}_i$ represents the theoretical values of the angular diameter depending on the mass $M$, the distance $D$ and the axion parameter $b$. $\mathcal{T}_i$ is evaluated assuming distances between $16.8\pm 0.8$ Mpc and masses in the range $M=6.2^{+1.1}_{-0.5}\times 10^{9}M_\odot$ and $M=3.5^{+0.9}_{-0.3}\times 10^{9}M_\odot$. The observed emission ring is expected to be 10\% larger than the true shadow size if the uncertainties related to the various accretion flow models are considered. Therefore, taking into account the $10\%$ offset, $\mathcal{O}=37.8\pm 2.7 \mu as$ also needs to be considered which is then compared with the model estimated values $\mathcal{T}_i$. 

\begin{figure}[t!]
\centering
\includegraphics[scale=0.75]{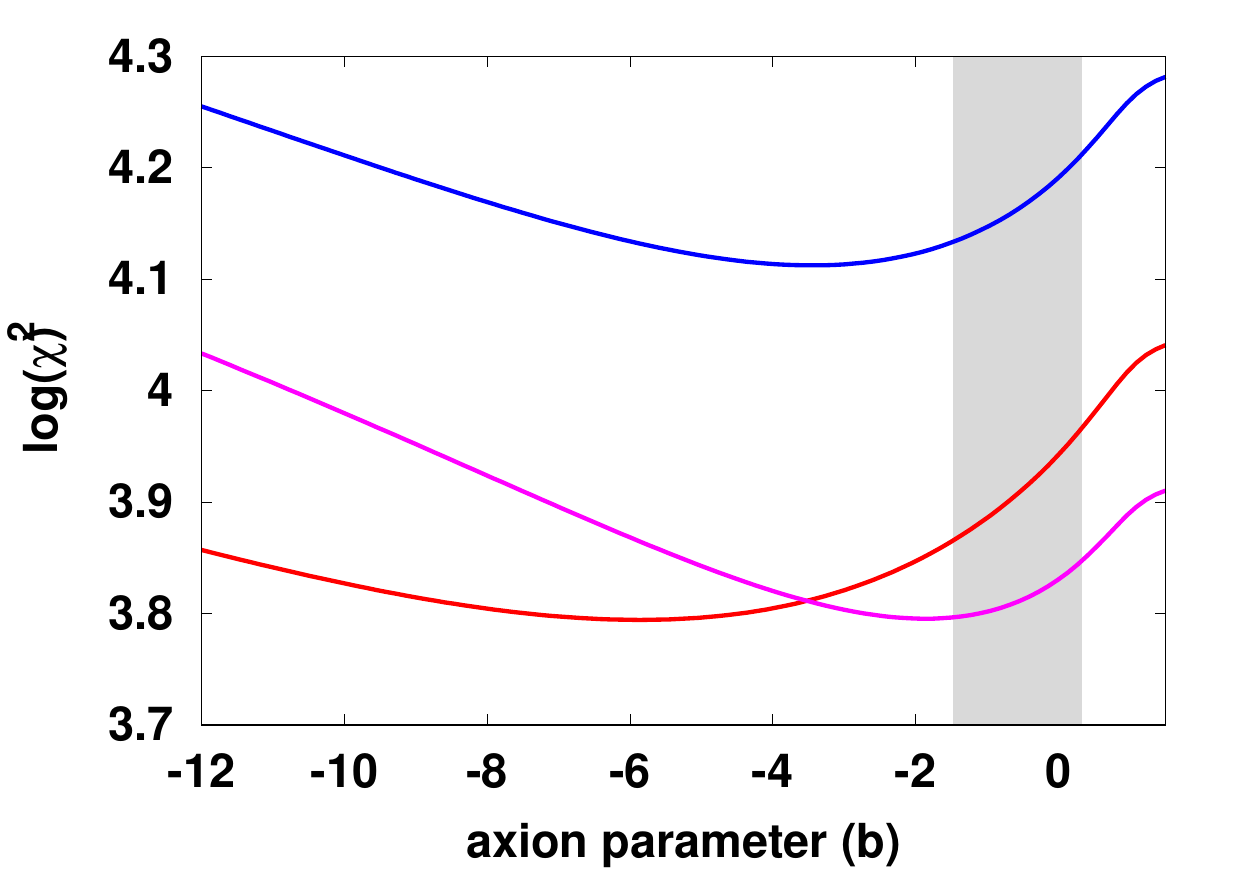}
\caption{The figure demonstrates variation of $\chi^{2}$ with the axion parameter $b$ obtained by considering the uncertainties in the distances, masses and the accretion flow models. The red and magenta curves correspond to the situation when $\mathcal{O}=42\pm 3 \mu as$ and $\mathcal{O}=37.8\pm 2.7 \mu as$ are used for comparison with the theoretical models. The blue curve on the other hand represents the scenario when the joint $\chi^{2}$ is computed taking into account both the aforesaid observations .
Interestingly, $\chi ^{2}$ attains the minimum value for a negative $b$, suggesting that the errors between theoretical estimates of angular diameter and the observations minimize when one considers axion with a negative energy density. }
\label{chi}
\end{figure}

For a given $b$, using these allowed values of masses and distances the $\chi^2$ is computed as in \ref{33}. The variation of $\chi^2$ with the axion parameter $b$ is plotted in \ref{chi} where the red curve corresponds to the situation when $\mathcal{T}_i$ is compared with $\mathcal{O}=42\pm 3 \mu as$ while the magenta curve represents the scenario when $\mathcal{O}=37.8\pm 2.7 \mu as$ is used to compute the $\chi^2$. The blue curve is associated with the joint $\chi^2$ when both $\mathcal{O}=42\pm 3 \mu as$ and $\mathcal{O}=37.8\pm 2.7 \mu as$ are compared with the theoretical values denoted by $\mathcal{T}_i$. All the three curves in \ref{chi} illustrate that $\chi^2$ attains the minimum for a negative value of $b$ where it is important to note that the signature of $b$ is crucial to this work while its exact magnitude is not so essential to achieve our conclusions. The shaded region in \ref{chi} represents the theoretically allowed range of $b$, $-1.48\lesssim b \lesssim 1/3$. In each of the three curves, the $\chi^2$ decreases monotonically as $b$ is reduced from $b_{max}=1/3$  to $b_{min}=-1.48$ which 
indicates that within the domain of the allowed values of $b$
a negative axionic parameter seems to explain the observation better. Such an axion violates the energy condition and has several interesting consequences which will be disussed in the next section. Interestingly, our present results are in concordance with a previous finding where we estimated the observationally favored signature of $b$ based on the spectral data of quasars. By comparing the observed spectrum of a set of eighty quasars with the theoretical spectrum from the surrounding accretion disk, we reported that the Kalb-Ramond field violating the energy condition (which is equivalent to a negative axion parameter $b$) seems to be favored by observations \cite{Banerjee:2017npv}. Since two independent astrophysical observations consistently favor a negative axion parameter it may be worthwhile to investigate this scenario in the context of other available observations, e.g. quasi-periodic oscillations in the power spectrum of black holes, implications on the signature of $b$ from gravitational wave observations.

%%%%%%%%%%%%%%%%%%%%%%

%%%%%%%%%%%%%%%%%%%%%%%
%%%%%%%%%%%%%%%%%%%%%%%%%%%%%%%%%%%%%%%%%%%%%%%%%%%%%%%%%%%%%%%%%%%%%%%%%%%%%%%%%%%%%%%%%%%%%%%%%%%
%%%%%%%%%%%%%%%%%%%%%%%%%%%%%%%%%%%%%%%%%%%%%%%%%%%%%%%%%%%%%%%%%%%%%%%%%%%%%%%%%%%%%%%%%%%%%%%%%%%
%%%%%%%%%%%%%%%%%%%%%%%%%%%%%%%%%%%%%%%%%%%%%%%%%%%%%%%%%%%%%%%%%%%%%%%%%%%%%%%%%%%%%%%%%%%%%%%%%%%%%%%%%%%%%%%%%%%%%%%%%%%%%%%%%%%%%%%%%%%%%%%%%
%%%%%%%%%%%%%%%%%%%%%%%%%%%%Appendix%%%%%%%%%%%%%%%%%%%%%%%%%%%%%%%%%%%%%%%%%%%%%%%%%%%%%%%%%%%%%%%%%%%%%%%%%%%%%%%%%%%%%%%%%%%%%%%%%%%%%%%%%%%%%%
\section{Summary}
\label{S5}
In this paper we aim to investigate the signatures of the Kalb Ramond field or its dual axion from the recent observations of the shadow of the supermassive black hole in the centre of the galaxy M87. This is important, since the weak field tests of gravity lack the necessary precision to discern the presence of such a field while the strong field tests e.g. electromagnetic spectrum emitted from the black hole accretion disk have reported that axion violating the energy condition is observationally favored.
Therefore, it is instructive to subject this finding to further tests and the observation of the black hole shadow provides the approprite opportunity.  

%Moreover, the observation of black hole shadow represents one of the 

%All these observational probes 

%solar system based tests of gravity as well the strong field tests involving electromagnetic emission from the black hole accretion disk have consistently  
In order to accomplish our goal, we compute the contour of the black hole shadow first in a general spherically symmetric background and note that the radius of the shadow depends only on the $g_{tt}$ component of the metric. Subsequently we consider the spherically symmetric solution of Einstein's equations solved in the Kalb-Ramond background. Such a metric exhibits a perturbation over the Schwarzschild scenario through the axion parameter $b$. Since the axion primarily appears as a $1/r^3$ correction to the $g_{tt}$ component of the Schwarzschild metric, its effect on the black hole image which probes the vicinity of the horizon, is expected to be significant.
Yet we can safely ignore the corrections to the metric with large inverse powers of $r$ due to the theoretical restriction on the axion parameter $-1.48\lesssim b \lesssim 1/3$, such that in our regime of interest $b/r^2$ continues to be less than unity. The fact that the magnitude of $b$ is very small from a theoretical consideration is further supported from the observations related to perihelion precession of mercury and bending of light \cite{Kar:2002xa}.
The theoretical lower bound on negative value of $b$ arises from the absence of any photon circular orbit outside the event horizon. This is an intriguing feature the spacetime exhibits due to the presence of the Kalb-Ramond field.

A stationary, axi-symmetric black hole solution of Einstein's equations minimally coupled to \KR field has not been obtained so far. This however, does not prevent us from constraining the axion parameter from shadow of M87* as the inclination angle of the object is very small $i\sim 17^\circ$. Consequently, even if M87* is a rapidly rotating black hole it will cast a circular shadow \cite{Cunha:2018acu,Vries_1999,Banerjee:2019nnj}. The choice of the spherically symmetric metric is therefore justified.
This is further supported by the fact that the observed shadow of M87* has an axis ratio $\Delta A<4/3$ and a deviation from circularity $\Delta C<10\%$ \cite{Akiyama:2019cqa}. Therefore, the angular diameter, $\Delta A$ or $\Delta C$ cannot be used to constrain the spin of M87*. We have verified this explicitly in a previous work \cite{Banerjee:2019nnj} where the tidal charge parameter of axisymmetric braneworld black holes could be constrained from the observed angular diameter, but nothing could be concluded about its spin from the observational constraint on $\Delta C$ and $\Delta A$. In this context we would like to mention that there exists the string inspired Einstein-Maxwell dilaton axion gravity where the stationary, axi-symmetric and asymptotically flat black hole solution has been worked out and is known as the Kerr-Sen solution in the literature \cite{Sen:1992ua,Ganguly:2014pwa}. The various fields associated with this theory, namely, the Maxwell field, the axion (or Kalb-Ramond) field and the dilaton field depend on  $r$ and $\theta$ and the solutions can be found in \cite{Ganguly:2014pwa}.   
The prospects of constraining such a metric from observations related to shadow of M87* has been discussed in \cite{Narang:2020bgo}.

We evaluate the dependence of the shadow radius on the axion parameter $b$ and find that
the radius of the shadow decreases with increase in $b$, or alternatively Schwarzschild metric perturbed with a negative axion parameter casts a larger shadow. It turns out that such an axion
or Kalb-Ramond field with a negative energy density bears a greater potential to reproduce the observed angular diameter of M87* assuming the known distances and masses of the object. We have shown that when the mass estimations of the object based on stellar dynamics ($M=6.2^{+1.1}_{-0.5}\times 10^{9}M_\odot$) or gas dynamics ($M=3.5^{+0.9}_{-0.3}\times 10^{9}M_\odot$) observations are considered, and the distance  obtained from stellar population measurements ($16.8\pm 0.8$ Mpc) are used to derive the angular diameter, a negative axion parameter turns out to be a better representation of the data.
Only when we consider $M\sim 6.5\pm 0.7 \times 10^9 M_\odot$ or allow a $10\%$ offset in the observed image, the $b=0$ model comes within the error bars. One however needs to note that the mass of M87* $M\sim 6.5\pm 0.7 \times 10^9 M_\odot$ is deduced from the angular diameter of its shadow assuming \gr\ and hence should not be ideally used to constrain the value of $b$ or other alternative gravity models from shadow related observations. 
Similarly the actual shadow size can be at most $10\%$ less than the observed emission ring (due to the uncertainties in the accretion processes), and we show that $b=0$ becomes viable only when this maximum offset is considered. 
To support our results, we perform a chi-square analysis taking into account all the uncertainties in the distances, mass and accretion models which explicitly reveals that for a negative axion parameter the $\chi^2$ attains a lower value.

The axion with a negative charge parameter has several interesting astrophysical and cosmological implications. It violates the energy condition and such
a scenario is often invoked for removal of singularity in geodesic congruences \cite{Kar:2006ms}, gains ground in bouncing cosmology to prevent the big bang singularity \cite{PhysRevD.77.044030}, plays a crucial role in altering the Buchdahl's limit for star formation \cite{Chakraborty:2017uku} and can potentially generate a non-zero cosmological constant in four dimensions whose origin is attributed to bulk Kalb-Ramond field in a higher dimensional scenario \cite{CHAKRABORTY2016258}.
Moreover, the suppression of Kalb-Ramond field has been discussed in several physical scenarios, e.g. in the context of warped brane-world models \cite{Randall:1999ee} with bulk Kalb-Ramond fields \cite{Mukhopadhyaya:2001fc,Mukhopadhyaya:2002jn} and the related stabilization of the modulus \cite{Das:2014asa}, in the context of higher curvature gravity where the associated scalar degrees of freedom diminishes the coupling of such a field with the Standard Model fermions \cite{Paul:2018ycm,Das:2018jey}, and in the inflationary era induced by higher curvature gravity \cite{Elizalde:2018rmz,Elizalde:2018now} and higher dimensions \cite{Paul:2018jpq}.  
%Suppression of Kalb-Ramond field strength has also been addressed previously in several physical scenarios, specially in the context of warped brane-world models \cite{Randall:1999ee} with bulk Kalb-Ramond fields \cite{Mukhopadhyaya:2001fc,Mukhopadhyaya:2002jn} and the associated stabilization of the modulus \cite{Das:2014asa}.

%The efficacy of such a scenario has been discussed in different context such as the bouncing model of the universe to avoid big bang singularity \cite{PhysRevD.77.044030}, the removal of singularity in geodesic congruences\cite{Kar:2006ms}, Buchdahl's limit for star formation \cite{Chakraborty:2017uku} and possible source of a spacetime with non-zero cosmological constant inherited from the bulk Kalb-Ramond field in a higher dimensional scenario \cite{CHAKRABORTY2016258}, where such energy violating term appears in an effective field theory due to the presence of an anti-symmetric tensor field.

As a final remark we mention that in the electromagnetic domain there is no dearth of spectral data of supermassive black holes while there is only a single observation of black hole shadow on which the present result is based. The real challenge of discering the nature of strong gravity from the black hole spectrum lies in appropriately modelling the spectrum which depends not only on the background spacetime but also on the nature of the accretion flow. Disentangling the impact of the metric from the spectrum therefore becomes quite non-trivial. The image of the black hole on the other hand provides a much cleaner environment to explore the strong gravity regime. However in this case, since the angular diameter is highly sensitive to the mass of the black hole, 
a precise measurement of the black hole mass is necessary to establish strong constraints on the signature of $b$. This in fact plays a crucial role in constraining $b$ rather than independent
observations of horizons with similar levels of uncertainties. In addition to this, the availability of more and more data on black hole images with reduced uncertainties, will further enhance the scope to constrain the signature of the axion parameter.

\section*{Acknowledgements}
The research of SSG is partially supported by the Science and Engineering
Research Board-Extra Mural Research Grant No. (EMR/2017/001372), Government of India.
The research of S.S is funded by CSIR, Govt.  of India.

\bibliography{Brane,KN-ED,Black_Hole_Shadow}

\bibliographystyle{./utphys1}
%\bibliographystyle{utphys1}
%\bibliography{x.bib}
%\bibliographystyle{plain}
%\bibliographystyle{JHEP}
%\bibliographystyle{plainnat}
\end{document}